# A Survey on Web Spam Detection Methods: Taxonomy

Shekoofeh Ghiam[1] and Alireza Nemaney Pour[2]

[1]Department of IT Engineering, Sharif University of Technology, International Campus, Kish Island, IRAN
`shokoofeh.ghiam@gmail.com`
[2]Dept. of Computer Software Technology Engineering,
Islamic Azad University of Abhar, Iran
`pour@abhariau.ac.ir`

## ABSTRACT

*Web spam refers to some techniques, which try to manipulate search engine ranking algorithms in order to raise web page position in search engine results. In the best case, spammers encourage viewers to visit their sites, and provide undeserved advertisement gains to the page owner. In the worst case, they use malicious contents in their pages and try to install malware on the victim's machine. Spammers use three kinds of spamming techniques to get higher score in ranking. These techniques are Link based techniques, hiding techniques and Content-based techniques. Existing spam pages cause distrust to search engine results. This not only wastes the time of visitors, but also wastes lots of search engine resources. Hence spam detection methods have been proposed as a solution for web spam in order to reduce negative effects of spam pages. Experimental results show that some of these techniques are working well and can find spam pages more accurate than the others. This paper classifies web spam techniques and the related detection methods.*

## 1. INTRODUCTION

Today lots of people have access to the Internet, and digital world has become one of the most important parts of every body's life. People not only use Internet for fun and entertainment, but also for business, banking, stock marketing, searching and so on. Hence, the usage of the Internet is growing rapidly. One of the threats for such technology is web spam. According to definition [1], a web spam is a web page which attracts search engine referrals (to this page or some other target pages) for the sole purpose. Suppose that you search a query in a popular search engine like Google but do not find relevant answer even from very first results. This is called web spam. Meanwhile it has a growing trend these days.

There are three different goals for uploading a spam page via spammers. The first is to attract viewers to visit their sites to enhance the score of the page in order to increase financial benefits for the site owners. The second goal is to encourage people to visit their sites in order to introduce their company and its products, and to persuade visitors to buy their productions. The last goal, which is the worst case, is to install malware on victim's computer.

Sometimes spammers steal critical information such as banking account and personal information by installing malware on the victim's computer. Spammers use different techniques to fool the search engines. As illustrated in Figure. 1, web spam techniques are classified into three categories as follows:

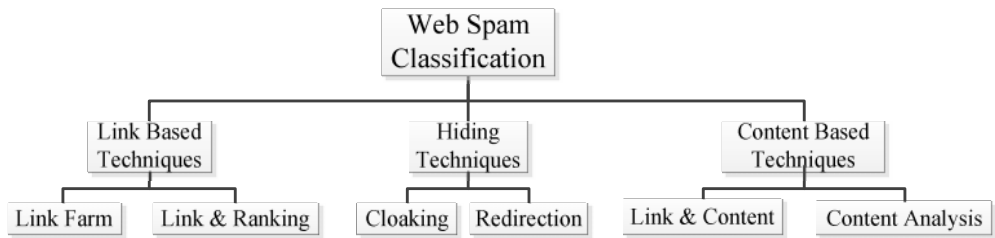

Figure 1. Classification of web spam techniques

(1) Link based techniques, also known as link spam [2]-[7]:the spammers try to add lots of links to target pages in order to manipulate search engine ranking algorithm. Sometimes spammers create a network of pages that are densely connected to each other, which is called link farms.

(2) Hiding technique [8]-[12]:this technique is divided to cloaking and redirection. Starting from the cloaking, spammers try to deliver two different contents to web crawler and normal visitors. A web crawler is a computer program that browses the World Wide Web in a methodical, automated manner. A web crawler has a database, which stores the record of requested pages. Hence, search engine ranking algorithms use web crawlers to evaluate web pages, so spammers try to deliver deceptive version of a URL to web crawlers to get higher score.

Redirection means the web page automatically redirects the browser to another URL as soon as the page is loaded. In this way, the search engine indexes the page, but the user does not notice it.

(3) Content-based technique [13]-[15]: the spammers try to retouch the content of target pages. For example a page may consist of all the words in a dictionary, so for an uncommon query this page will be chosen with high score, or will be the only result. Another trick is repeating popular words many times in the target pages in order to increase page's score in search engine ranking.

Detecting spam pages is important because of three main reasons. First, many spam pages are harmful for both search engines and victim's machines. Second, spam pages waste visitors' time because there is no useful information in such pages. In addition, this may cause some kinds of distrust to search engine results. Finally, spam pages waste important resources of search engines such as network bandwidth at crawling, CPU cycles at processing and storage space while indexing. Since it is very hard and time consuming for human experts to check all pages manually and determine whether the pages are spam or not, it is required to find suitable ways to detect web spam more quickly and accurately.

The proposed solution for web spam technique is the detection method. There are some detection methods [2]-[15] in regard with the web spam techniques. We will discuss these methods in section 2. The rest of this paper is organized as follows. Section 2 discusses web spam techniques and the related detection methods. Section 3 compares the existing methods briefly. Finally, the conclusion is given in section 4.

## 2. WEB SPAM DETECTION METHODS

This section discusses web spam detection methods based on its classified techniques shown in Figure 1. As stated before, the web spam techniques used by spammers are classified into three big categories; which are link based, hiding and content-based techniques. Based on these techniques, there are different web spam detection methods [2]-[15]. We will discuss these detection methods in details in the following subsections.

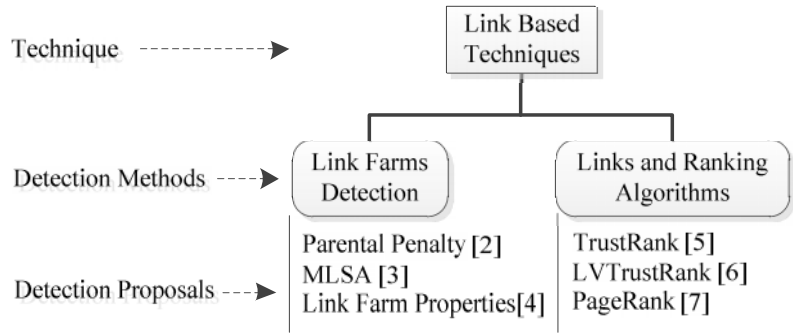

Figure 2. Detection methods and the related proposals for link based technique

## 2.1. Detection Methods for Link Based Techniques

The detection methods proposed for link-based technique are classified into link farm detection method [2]-[4], and links and ranking algorithms method [5]-[7]. Figure 2 illustrates the classification of the detection methods and the related proposals for link based techniques.

### 2.1.1. Link Farm Detection Methods

Link farm spam refers to a technique which tries to create a network of pages densely pointed to each other. In this way, the incoming and outgoing links of a web page increase rapidly. Consequently, search engine ranking algorithms may give a high score to such pages because of the large number of links. However, the content of such pages may be very far or completely irrelevant to the searched query.

### A)  *Parental Penalty*

Wu and Davison [2] proposed a detection method to find link farms. This method consists of two steps. The motivation behind this method is that pages in a link farms are densely connected to each other, so pages within link farms usually have several common nodes between the incoming links set and outgoing links set. If we find a seed set, for a new page, it is possible that the page is a part of a same link farm. Then we can expand this seed set to find other link farms. To explain this method briefly, we can say that it first generates a seed set and next expands it to find link farms. For expansion step they used a method called Parental Penalty which is the name of the whole algorithm too. This method uses a threshold in order to increase the accuracy. For evaluation the threshold is set to 3.

Figure 3 illustrates a simple example of a network of pages, which creates a link farm. In this example, each node denotes a web page with a threshold set to 2. As shown, the incoming link set for node A is [C, D, E], and the outgoing link set is [B, C, D]. The common nodes between the incoming and outgoing link sets are [C, D] which is equal to the predefined threshold, so node A is put to the seed set, and it has a very high possibility to be a spam page. Similar to node A, nodes C and D are put to the seed set because the number of common nodes in the incoming and outgoing link sets are equal or above the threshold. Finally the seed set is [A, C, D].

The next step is the expansion step. For this step, they used a method called Parental Penalty. This method is used to penalize pages that point to bad pages. They assume that a page pointing to a bunch of bad pages is likely a bad page. The threshold is set again to 2. In Figure 3, node E has two outgoing links to node A and D, which are in the seed set from the previous step. Node

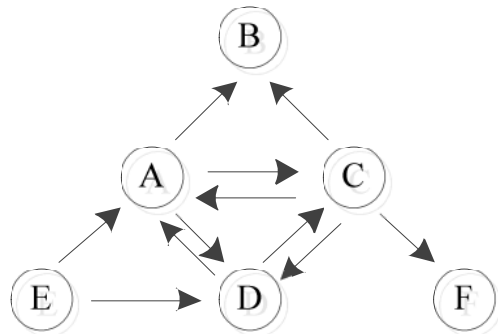

Figure 3. A simple example of link farm with 6 pages [2]

B has one link, and node F has zero links to the seed set. Based on the threshold, node E is added to the seed set.

For the evaluation of this method, 200 queries were selected. For each query, the top 10 URLs generated from Kleinberg's HITS and their "weighted popularity" after dropping links among spam pages are mixed together for blind evaluation. The advantage of this evaluation is that domain matching was used instead of only page matching. After the evaluation, it was proven that the precision of this method is high. Precision is defined as the proportion of selected items that the system gets right. Higher precision indicates that the algorithm is better. However, weaknesses have been reported for this method. By this method, duplicate pages cannot be detected, and consequently this is a high possibility of false negative. It means that the algorithm incorrectly reports that spam pages were not detected while such pages were really present.

### B) MLSA

To overcome the false negative problem mentioned above, Tung and Adnan [3] proposed an improved version of the previous algorithm. This method is called Multi level Link Structure Analysis (MLSA). This method focuses more on the link exchange not only between the pages in the same domain, but also between pages in different domains. This method is based on the idea that in most link farms, pages in the same domain have at least one outgoing link from one of the pages in the domain to the neighbouring domain. Figure 4 illustrates a simple example of MLSA algorithm. Pages $A_1$, $A_2$, $A_3$ and $A_4$ are in the same domain. Page $A_1$ is selected as the candidate page for analysis. In the first step of the algorithm, the outgoing links of page $A_1$ to the pages in the same domain and other domains is collected. $A_2$ and $A_3$ are the pages that are going to be analysed in the next step because $A_1$ points to them and also they are in the same domain with $A_1$. Next, the outgoing links of $A_2$ and $A_3$ to the pages in the same domain and other domains are collected. The algorithm continues this procedure depending on precision required by the experiment. Finally, the algorithm counts the common domain names of all incoming and outgoing links. If this number is above the predefined threshold, this page will be marked as a bad page.

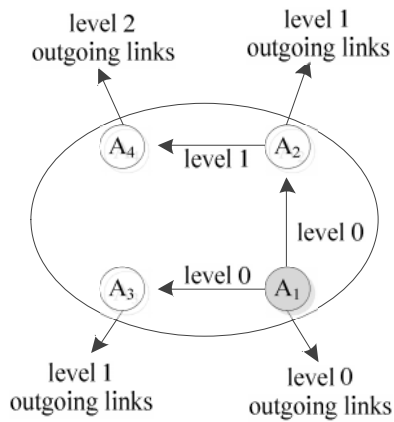

Figure 4. Link parsing sequence of MLSA [3]

Experimental results using Yahoo! Search Engine show that MLSA can find more spam pages compared to previously proposed methods. However, there is a potential for false positive. False positive occurs when the algorithm incorrectly reports a page as a spam page, when it was really not spam. It seems that one improvement will be achieved by adding the content relevancy property of a web page to MLSA algorithm.

### C) Link Farm Properties

Wang et al.[4] proposed a detection method that uses some properties of a network of pages to handle web spam. If we consider a World Wide Web as a graph, each page is considered as a node, and each link is considered as an edge. The degree distribution and average path length of websites are two properties. Degree distribution property is the probability of a randomly selected vertex with exactly $K$ edges. The average path length of a network is a property which the sized of a network is measured. This property uses the number of edges in the shortest path and number of vertices in the network.

There are two kinds of networks, a scale free network, and a slightly fully connected network. A Scale free network is a network whose degree distribution has low distribution power. Example of scale free network is World Wide Web. In slightly fully connected networks a node is not connected to all other nodes in the network compared with fully connected networks that each node has links to all other nodes. Figure 5 shows the modes of network.

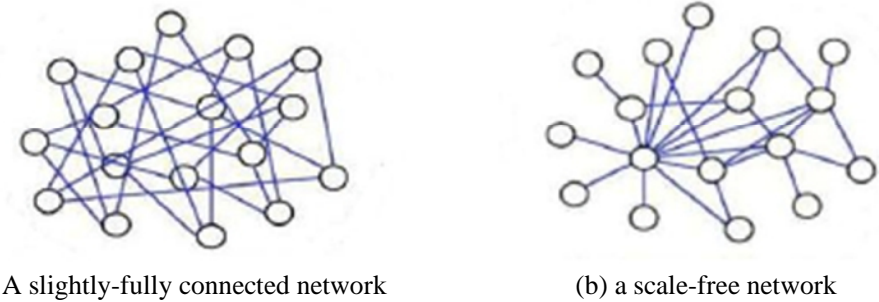

(a) A slightly-fully connected network        (b) a scale-free network

Figure 5. The modes of network [4]

This detection method consists of two approaches. The main idea behind the first approach is that if the degree distribution of a tested website follows irregular distribution, the decision with confidence will be that this web site is using link farm spam. The second approach notices the average path length property. For a non-spam web page the value of this property should be near the value calculated by $L=0.35+2.06\ log\ (N)$, where $N$ is the number of vertex in the network. If the value of average path length is smaller than the one computed by this formula, there is a high possibility that this page is a part of link farm.

**Evaluation:** Experimental results for both approaches show that this method can find link farms by using whole link structure of web site.

*D) Comparison between Link Farms Detection Methods*

Each of these methods has its own pros and cones. However, the last method, Link Farm Properties, is the improved version of the previous ones. Since this method finds spam pages based on mathematical evaluations, it can find web spam more accurate than other methods. In addition, the possibility of false negative and false positive is low compared to previous methods.

### 2.1.2. Links and Ranking Algorithms

There is an old method called PageRank. The motivation behind this method is that the importance of a page is being influenced by the importance of some other pages. . It means that a web page is important if several other important web pages point to it. The PageRank score for a web page is computed with a formula.

*A) TrustRank*

Gyöngyi et al. [5] proposed an algorithm called TrustRank. This detection method semi-automatically finds good pages among lots of pages. Since finding spam pages completely manual by human experts is very hard and time-consuming, this algorithm tries to find good pages semi-automatically. The main idea of TrustRank algorithm is that good sites rarely point to spam sites. There are two important parts in this algorithm: the first part is selecting a seed set, and the second part is finding good pages.

Selecting seed set: the goal of selecting seed set is to find pages that will help us more to find additional good pages. This method introduces two ways to find seed set. The first approach is called "Inverse PageRank". In this approach, pages will be selected for seed set based on the number of outlinks. It means that pages with higher number of outlinks will be in seed set. This approach is the inverse version of PageRank. Since in PageRank, pages with higher number of in links are selected.

Another way to find seed set is "High PageRank". This method finds seed set based on the higher PageRank of web pages. Since high PageRank pages are likely to point other high PageRank pages, relevant results will be in top of the results of search a query. While there is higher trust to seed set created by High PageRank, a disadvantage is appeared. We may identify the goodness of fewer pages in comparison with Inverse PageRank. However, these pages are more importantand more reliable than pages in seed set created by Inverse PageRank. Experimental results show that this algorithm can filter out spam pages more effectively than PageRank.

*B) LVTrustRank*

Chen et al. [6] discovered that there are many spam pages that have many incoming links from good pages, so they concluded that TrustRank algorithm is invalid in such cases. This method notices the variance of link structure in order to improve TrustRank algorithm. The new algorithm is called LVTrustRank. The formulas used in LVTrustRank are as follows:

$$t_f = (\frac{t_1^* + t_2^*}{2})^{1+IGR}$$

$$IGR = \frac{\left|S_{in}(t_1)\right| - \left|S_{in}(t_0) \bigcap S_{in}(t_1)\right|}{\left|S_{in}(t_0)\right|}$$

(1)

Refer to (1), $t_1^*$ denotes TrustRank score at time $t_1$ and $t_2^*$ is the TrustRank score at time $t_2$. *IGR* stands for inlink Growth Rate, which is defined as the ratio of the increased number of inlinks at the site to the number of original inlinks. For evaluation, some pages with more outgoing links were selected. Then, they were put together to create a seed set. For each page in the seed set, the algorithm computes *IGR* value, and removes pages with high *IGR* value.

Experimental results show that LVTrustRank can find a few more pages compared with TrustRank. However, LVTrustRank does not work always better than TrustRank because some spam pages do not often change their link structure, and consequently finding these kinds of spam pages by LVTrustRank is very hard.

### C) PageRank

As we mentioned earlier, PageRank is a link structure-based algorithm. In PageRank algorithm the PageRank value (PR) for a page is divided between its entire outlinks. In addition, highly linked pages are more important than pages with few links. However, there is an obvious weakness in this algorithm because some pages are more important than others and giving equal values to all outlinks is not fair. Therefore, this algorithm is blind to find spam pages correctly.

In order to handle the stated issue, Pu et al.[7] proposed a new PageRank algorithm introduces the idea of popularity of web pages. The main idea of this method is that although pages with higher number of outlinks are more important than others, they are not always popular. The new PageRank algorithm divides the *PR* score between outlinks based on the weight of each outlink. In addition, when a page is outlinked from one important page, high PageRank is assigned. Meanwhile, the less important pages get low PageRank. Based on these assumptions, new formulas are derived. Experimental results show that the old PageRank algorithm has been improved, and the new algorithm can effectively find more spam pages.

### D) Comparison between Links and Ranking Algorithms

Experimental results show that the new PageRank algorithm works better than the other algorithms in this category. As the algorithm is weighted, it is fairer and more accurate than the previously proposed methods.

As we mentioned earlier, in link based spamming techniques, there are two spamming methods. The first one is creating link farm, and the second one is that a good page is linking to spamming page. Between these two methods the latter is harder to defend because it uses a kind of deception which makes it harder to detect such pages compared to the former method. However, link farms may harm search engines more. Since they create a network of pages densely connected to each other, search engines may evaluate these pages as useful information about the searched query with high probability and put them in the first results of searching while in reality it is not true, and these pages may have no useful information.

## 2.2. Detection Methods for Hiding Techniques

This section provides web spam detection methods and related proposals for hiding techniques in details [8]-[12]. Generally, hiding technique consists of cloaking and redirection techniques. The detection methods are defined as cloaking detection methods [8]-[12], and redirection detection method [9]. Figure 6 illustrates the classification of detection methods and the related proposals for hiding technique.

## 2.2.1. Cloaking Detection Methods

Cloaking occurs when different copies of a URL are sent to web crawlers and web browsers. Therefore, the perspectives of crawler and browser become different. For example, when a URL is cloaked in any way, it is intended for raising the ranking of specific URLs in the search engine results not intended for human viewing. However, the content received by a web browser may have a normal-looking, human-readable web page or simply a broken link such as "HTTP 404: file not found".

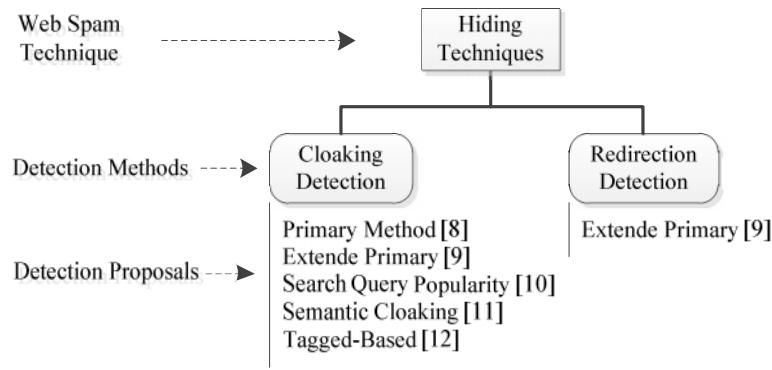

Figure 6.Detection methods and the related proposals for hiding technique

There are two different cloaking methods; syntactic cloaking and semantic cloaking. Syntactic cloaking is simpler than semantic cloaking, and is referred to the case of whether web crawlers and web browsers have received different content. On the other hand, semantic cloaking is referred to differences in the content of web crawlers and web browsers intended to deceive search engine ranking algorithms.

## A) *Primary Method*

Najork [8] proposed a primary method, which compares only two copies of a URL, crawler and browser's copies. Let assume $C_i$ and $B_i$ are copies of the crawler and the browser, respectively. If the crawler and the browser perspective is completely the same, this page is not considered a cloaked page. Otherwise, it is classified as a cloaked page. The weakness of the proposed method is that the dynamically generated or frequently updated pages are classified as cloaked pages falsely. In order to improve its accuracy, further copies of a URL is needed.

Wu and Davison[9] proposed algorithms, which were more accurate than the previous works. In fact, it extends the primary method proposed by Najork [8]. First, two different methods, Term differences and Link differences were introduced to detect cloaking. Starting from term differences method, the algorithm calculates the number of different terms between $C_1$ and $C_2$, called NCC, then, it calculates the number of different terms between $C_1$ and $B_1$, called NBC. In addition, a threshold is predefined. If NBC > NCC, this page has a high probability to be a cloaked page. In link differences method, the algorithm calculates the number of different links between $C_1$ and $C_2$, called LCC. Then, the algorithm calculates the number of different links between $C_1$ and $B_1$, called LBC. By using a predefined threshold, if LBC > LCC, then there is a high probability that this page is a cloaked page.

In addition to above algorithm [9], a new algorithm was proposed to detect syntactic cloaking automatically. In this algorithm, four copies of a URL, two from the crawler's perspective and two from the browser's perspective are needed. The algorithm calculates the number of common terms within $B_1$ and $B_2$ which does not exist in $C_1$ or $C_2$, called TBNC. Moreover, the algorithm calculates the number of common terms in $C_1$ and $C_2$ which does not exist in $B_1$ or $B_2$, called TCNB. When TBNC+TCNB get greater than the predefined threshold, this page is classified as a cloaked page.

Experimental results with two data sets, HITSdata and HOTdata for these extended algorithms show that cloaking occurs in 3% of the first dataset and 9% of the second dataset. However, the higher threshold gives higher precision and lower recall. Therefore, there is a kind of trade off between precision and recall. Hence, finding satisfactory threshold is important.

### B) Search Query Popularity

Chellapilla et al. [10] proposed an algorithm which calculates cloaking score for each requested URL based on popularity and monetizability. This algorithm uses normalized term frequency difference. In fact, it checks differences between two copies of a URL sent to web crawler and web browser.

The proposed algorithm proposed works as follows. At first, it retrieves $C_1$ and $B_1$. If they are from the same HTML pages, it concludes that there is no cloaking in this page. If they are from different HTML pages, those pages are converted to text. If the texts are equal, it concludes that there is no cloaking in this page. Otherwise, term differences between $C_1$ and $B_1$ are calculated. If the result is zero, it concludes that there is no cloaking in this page. If the result is not zero, $C_2$ and $B_2$ are retrieved. Then, cloaking score ($S$) is calculated by using (2). If $S$ is greater than the predefined threshold, this page is classified as a cloaked page. If not, it is a normal page.

$$S = \frac{\Delta D}{\Delta S}$$

$$\Delta S = \max(D(C_1, C_2), D(B_1, B_2)) \tag{2}$$

$$\Delta D = \min(D(C_1, B_1), D(C_2, B_2))$$

For evaluation, two lists of 5000 queries were used for data set. The first list was the set of the top 5000 most popular search queries computed over one month. The second list was the set of the top 5000 most monetizable search queries over one day. For each query, the top 200 search results were obtained from three search engines, Google, MSN, and Ask. There were about 1.49 million unique URLs in popular set and about 1.28 million unique URLs in monetizable set. Experimental results show that this algorithm has a high accuracy in detecting spam pages in monetizable query results.

### C) Semantic Cloaking

Wu and Davison [11] proposed a two-step cloaking detection method which detects semantic cloaking on the web. The first step is called filtering step. The motivation of creating this step is to find uncloaked pages (normal pages) by less cost. In this step, $C_1$ and $B_1$ are compared. To avoid classifying dynamically generated or frequently updated pages as cloaked, a satisfactory threshold is used. It means that if the term difference between $C_1$ and $B_1$ is greater than the predefined threshold, then this page will be sent to the next step of the algorithm. Otherwise the algorithm concludes that this page is a normal one. Therefore, the algorithm can find uncloaked pages only by retrieving two copies of a URL instead of four copies. Hence, cost will be reduced.

Candidate pages from this step will be sent to the second step of the algorithm. In the second step, called classification step, $C_2$ and $B_2$ are retrieved and a set of features is generated based on $C_1$, $C_2$, $B_1$ and $B_2$. This set of features includes number of terms, number of links, common terms, number of terms in meta-keyword, the ratio of number of related links to the total number of links, etc. Finally, for identifying cloaking, a $C_{4.5}$ decision tree based on these features is generated.

For evaluation, ODP data set was collected. The 2004 ODP RDF file were used, and 4,378,870 URLs were extracted from it. Experimental results show that this algorithm has a precision of 93%. The recall of this algorithm is 85%. Recall is calculated as the percentage of cloaked URLs in the data set that are correctly identified.

### D) Tagged-Based

Jun-Lin[12] proposed tag-based method to detect cloaking. The motivation behind this method is that tags in a web page do not change as much as links and terms in a web page. Therefore, tag based method detects cloaking better than previous methods. Three tag-based cloaking detection methods are proposed. In the first one, two copies of a URL, $C_1$ and $B_1$, are retrieved. In the second method, three copies of a URL, $C_1$, $B_1$ and $B_2$ are retrieved. In the third method, $C_1$, $B_1$, $C_2$ and $B_2$ are retrieved. Based on union, intersection and difference, these three algorithms are called $TagDiff2$, $TagDiff3$ and $TagDiff4$ respectively.

$$TagDiff2 = \left| B_1' \setminus C_1' \right| + \left| C_1' \setminus B_1' \right|$$

$$TagDiff3 = \left| B_1' \setminus C_1' \right| + \left| C_1' \setminus B_1' \right| - \left( \left| C_1' \setminus C_2' \right| + \left| C_1' \setminus C_2' \right| + \left| C_2' \setminus C_1' \right| \right) \qquad (3)$$

$$TagDiff4 : \left| (B_1' \cap B_2') \setminus (C_1' \cup C_2') \right| + \left| (C_1' \cap C_2') \setminus (B_1' \cup B_2') \right|$$

For evaluation, four sets of queries were selected from the web in order to generate data set. The first set(denoted as WQ) contained weekly popular queries of Google from 1/1/2006 to 23/12/2006. This set comprised 710 queries. It was reduced to 595 queries after removing the duplicates. The second set (denoted as MQ) contained monthly popular queries in 2006 from both Google and AOL. This set comprised 775 queries (665 from Google and 110from AOL). It was reduced to 484 queries after removing duplicates. The third set (denoted as YQ) contained the yearly popular queries from six sources, Google, Yahoo, MSN,AOL, Ask, and Lycosin 2006. The fourth set (denoted as DQ) comprised the directory names from the top two levels of ODP,but excluding the descendants of the ''World'' directory to avoid adding 79 language names to DQ. Every query in these four sets was entered into Google, and its top 200 search results were collected. Finally, 1% of URLs were randomly sampled, and then manually were checked whether each URL was cloaked in the sample. Experimental results show that there is an enhancement in both precision and recall compared to term based and link based methods. Therefore, this method is more accurate than previous ones.

### E) Comparison between Cloaking Detection Methods

It is obvious that tag-based method is working better than previously proposed ones. This method increases precision while recall is reasonable too. However, since spammers try to change the content of web pages in order to manipulate search engine ranking algorithms, Extended Primary [9] and Semantic Cloaking [11] which work on the content of web pages will give reasonable results in long-term.

Table 1. Number of pages using different types of redirections [9]

| Type | Crawler | Browser |
| --- | --- | --- |
| 301 | 20 | 22 |
| 302 | 50 | 60 |
| Refresh Tag | 4230 | 4356 |
| JavaScript | 2399 | 2469 |

### 2.2.2. Redirection Detection Method

Redirection is defined as redirecting the browser to another URL automatically as soon as the page is loaded. In this way, the page is still gets indexed by the search engine, but the user does not notice it. Wu and Davison [9] evaluated redirection occurring. Results show that about 153 pair of pages out of 250000 pair has different response codes to web crawler and web browser. For example, the web sites sent the response code 404 or 503 to one and the response code 200 to the other. Table 1 shows number of pages using different types of redirection.

As we mentioned earlier, hiding techniques are divided into two categories; cloaking and redirection. Between these methods, cloaking techniques are harder to defend because spammers send two different versions of a web page to both web crawler and web browser. For this reason, search engine ranking algorithm may have difficulty to identify whether this page is cloaked or not. In addition, cloaking is more popular between spammers, and it is used more than redirection one. Therefore, cloaking method is harder to defend compared to redirection one.

### 2.3. Detection Methods for Content-BasedTechniques

In this section, we discuss content-based detection methods and the related proposals for content-based techniques as shown in Figure 7. The detection methods consist of two parts. The first method uses the combination of link and content to detect web spam while the second method utilizes the features of the web page to find manipulated contents.

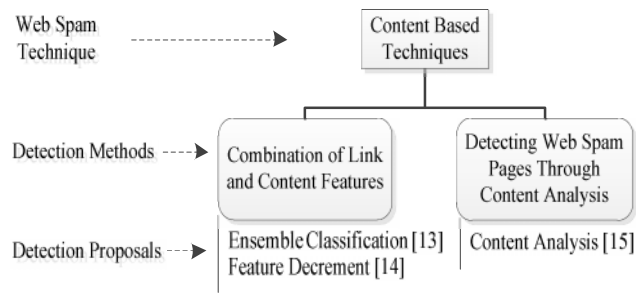

Figure 7.Detection methods and related proposals for content-based techniques

### 2.3.1. Combination of Link and Content Features

#### A) Ensemble Classification

Geng et al. [13] proposed a method which used spamicity ensemble undersampling classification strategy. The spamicity is defined as the probability that a website belongs to web spam. This probability is shown *PS*. The motivation behind this method is that in reality the ratio of spam websites is lower than reputable websites. Therefore, gathering spam pages is more difficult than gathering reputable pages. The spamicity is calculated by (4) where *x* is a test sample, $C$ shows a specific classifier, and $P_{spam}\ (x,C)$ and $P_{normal}\ (x,C)$ are predicted probability of $x$ belonging to spam set or not respectively with classifier $C$. PS is the prediction spamicity which is the probability of a website belonging to web spam. According to the distance or posterior probability of a sample, PS can be shown as follows:

$$PS(x,C) = \frac{P_{spam}(x,C)}{P_{spam}(x,C) + P_{normal}(x,C)} \tag{4}$$

For the experiment, three different classifiers, $C_{4.5}$, Bagging, and Adaboost were used. By using spamicity formula (4) and these classifiers, an algorithm was proposed. This is the first proposal which used both the content and link based features to detect web spam pages. Using WebSpam UK-2006 dataset, experimental results show that F_measure, which is the performance measurement based on precision and recall, was improved. Among the classifiers, Adaboost showed the best results. However, it's better to evaluate this method using more learning algorithms.

### B) Feature Decrement

Mahmoudi et al. [14], proposed a method for web spam detection based on discriminative content and link features. The main idea of this proposal is to reduce the number of features inorder to increase the performance of classifier while the accuracy is kept in the same level, or even it is improved. Therefore, the irrelevant features are filtered out.

In this method, the correlation between features is an important factor. The results show that this correlation is very effective in web spam detection. Hence, 10 features are selected as the final features. For classification, different algorithms such as neural network, SVM, Naïve Bayes and Decision trees were used. Finally LDAtree was selected because it worked better than the other algorithms. Four performance measurements, precision, recall, F-measure and ROC were used to evaluate this idea. ROC is a plot of true positive rate versus false positive rate. Using WebSpamUK 2007,experimental results show that there is not only no reduction in the performance of web spam detection, but also an improvement has been achieved. Results are shown in Tables 2,3 and 4.

Table 2.The results of experiment without feature selection methods [14]

| Algorithm | # of Features | Precision | Recall | F_Measure | ROC |
|---|---|---|---|---|---|
| LADtree | 140 | 66.7% | 66.6% | 66.6% | 73.4% |
| Neural Network | 140 | 67.3% | 67.3% | 67.3% | 72.7% |
| SVM | 140 | 70.6% | 69.6% | 69.3% | 69.6% |
| Naïve Bayes | 140 | 63.4% | 62.8% | 62.3% | 67.5% |

Table 3.The ROC comparison of the results experiments with and without feature selection [14]

| The ROC Comparison | # of Features | LADtree | Neural Network | SVM | Naïve Bayes | Random Forest |
|---|---|---|---|---|---|---|
| All Features | 140 | 73.4% | 72.7% | 69.6% | 67.5% | 76.3% |
| 2 | 118 | 72.1% | 70.6% | 64.1% | 71.4% | 74.1% |
| SVM | 93 | 72.5% | 65.9% | 70.7% | 68.9% | 75.4% |
| IG | 53 | 74.5% | 69.1% | 63.3% | 64.4% | 74.6% |
| CFS | 26 | 76.2% | 70.6% | 64.1% | 71.4% | 74.1% |

Table 4. The F-measure comparison of the results experiment with and without feature selection [14]

| The F-meaure Comparison | # of Features | LADtree | Neural Network | SVM | Naïve Bayes | Random Forest |
|---|---|---|---|---|---|---|
| All Features | 140 | 66.6% | 67.3% | 0.69.3% | 0.62.3% | 0.71.4% |
| 2 | 118 | 66.6% | 57.7% | 68.4% | 63.6% | 0.69.3% |
| SVM | 93 | 68.8% | 68.4% | 70.3% | 68.9% | 69.2% |
| IG | 53 | 68.8% | 65.5% | 63.1% | 43.2% | 69% |
| CFS | 26 | 70.7% | 65.1% | 63.8% | 65.3% | 66.4% |

### C) Comparison between DetectionMethods in this Category

One of the most important features in detection methods is reducing time and cost. The algorithm proposed in [14] achieves these goals. This algorithm not only reduces the irrelevant features, but also improves the performance. Therefore, this detection method works better than the other method.

## 2.3.2. Detecting Web Spam Pages through Content Analysis

### A) Content Analysis

Ntoulas et al. [15] explored a variety of methods for detecting spam. At first, they tried to check whether some pages are more likely to be spam or not. This experiment contains two parts. In the first part, some special top-level domains were checked to find out whether the kind of the domain is important for a page to be a spam page or not. Results show that for example in .biz domain about 70% of all pages are spam, and no spam page was found in .edu domain.

In the second part, the language of pages was checked to find out whether the language of a web page has any effect on being spam or not. Results show that about 25% of French pages are most likely to be spam while only 4.5% of Chinese pages are spam. In addition, other features of a web page were checked to find their effectiveness in spamicity. Number of words in the page was one of those features. Results show that the pages with more words are likely to be spam. However, because of the probability of false positive, we cannotstrictly classify a web page as a spam page by only counting the number of words.

Another feature is the amount of anchor text in a web page. There are spam pages, which solely provide anchor text for other pages. Results show that higher number of anchor text in a web page may increase the probability of spam for that page. Like the previous feature, the probability of false positive exists when we only use anchor text to find spam pages.

The next feature which was checked is the fraction of visible content. Some information may not visible for web browsers. This information may be only used for manipulating search engine ranking algorithms. Results show that most of spam pages contain fewer markups compared to normal web pages.

Compressibility is another feature. By compression, we have not only reduction in the space and disk time, but also in the redundancy since a compressor can represent a second copy using as a reference to the first one. For evaluation, compression ratio, the size of the uncompressed page divided by the size of the compressed page, was used. Results show that pages with the compression ratio of at least 4 are more likely to be spam.

Finally, it is important to know that using each heuristic alone will not give an accurate web spam detection result. The main idea is to combine the outcome of every heuristic to have more accurate method. By considering spam detection problem as a classification problem, the heuristic combination will be provided. Using C4.5 classifier for evaluation, experimental results show that by using this detection method almost 86.2% of all spam pages are detected correctly.

In order to improve the accuracy of the classifier, various techniques are available. *Bagging* and *boosting* are the two most popular ones. Both of these techniques essentially create a set of classifiers, which are then combined to form a composite classifier. Results are shown in Tables 5, 6 and 7.

Table 5. Recall and precision of the classifier [15]

| Class | Recall | Precision |
|-------|--------|-----------|
| Spam | 82.1% | 84.2% |
| Non-Spam | 97.5% | 97.1% |

Table 6. Recall and precision after bagging [15]

| Class | Recall | Precision |
|-------|--------|-----------|
| Spam | 84.4% | 91.2% |
| Non-Spam | 98.7% | 97.5% |

Table 7. Recall and precision after boosting [15]

| Class | Recall | Precision |
|-------|--------|-----------|
| Spam | 86.2% | 91.1% |
| Non-Spam | 98.7% | 97.8% |

In this section, there are two possible spamming techniques, combination of link and content features and manipulating the content of web pages. Between these two techniques, the second method, manipulating the content of web pages, is harder to defend since spammers use different techniques to manipulate the content to make it difficult to defend. For example, they use special domains, special languages, or anchor text etc. Also they use invisible contents in web pages or repeat some key words many times. Recognizing these kinds of trickery is harder than the first content based spamming technique. In addition, this technique has the most destructive effect on search engines. As we mentioned earlier, spammers use different ways to fool search engine ranking algorithm, the problem will be harder when they use the combination of tricky ways. Therefore, the destructive effect of this method is more than the first one.

## 3. FINAL COMPARISON

Each of the explained methods uses a kind of trickery to fool search engine ranking algorithms. In link-based techniques, spammers try to fool search engine ranking algorithms by adding many links to special pages, or try to have links from good pages. In hiding techniques, spammers deliver different contents to web crawler and normal visitors, so there is a difficulty to recognize this kind of spamming technique. The last method is using the features of web pages to fool search engines. They use different trickery ways to add special key words or terms to web pages. Between these methods, hiding techniques are harder to defend. Since spammers deliver different contents to web crawlers and web browsers, detecting this kind of cloaking needs comparison between the content of the page sent to web crawler and normal visitor. This comparison is time consuming. Also there is no automatic method in this area to find cloaking or redirection.

Content-based techniques have the most destructive effect on search engines. Since spammers can easily add key words or special terms visibly or invisibly to web pages, this method is a popular one between spammers. In addition, if we search a query in a popular search engine like Google, the first spamming technique that we encounter will be content-based techniques.

According to experimental results achieved from different proposed methods in different areas, methods proposed in hiding techniques can find spam pages better than other methods. These methods can find more spam pages with high precision and reasonable recall. Consequently, these methods will help search engines more.

## 4. CONCLUSION

We talked about three important web spam techniques used by spammers to manipulate search engine ranking algorithms. Among these techniques, there are few papers available for cloaking and redirection. Also there is high potential for addressing content-based techniques. As a solution for web spam issue, web spam detection method has been proposed. These methods will help search engine ranking algorithms allocate ranking score more accurate than before. Therefore, spam pages will get lower score. We believe that there is still high need for researching in this area to help search engine ranking algorithms to find spam pages.

## *Authors*

**Shekoofeh Ghiam** has received her B.Sc. from Islamic Azad University in software engineering and the M.Sc. in Information Technology engineering from Sharif University of Technology, International Campus, Iran in 2011. Her research area is about Internet Security, Wireless Sensor Networks and its application.

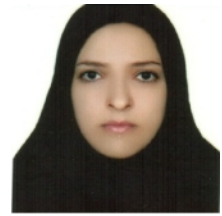

**Alireza Nemaney Pour[1]** has received his B.S degree in Computer Science from Sanno University, Japan, M.S in Computer Science from Japan Advanced Institute of Science And Technology, Japan, and Ph.D. degree in Information Network Science from Graduate School of Information Systems, the University of Electro-Communications in Japan.

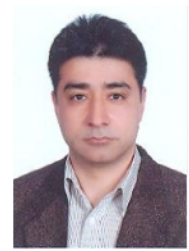

He is currently a faculty member of Islamic Azad University of Abhar in Iran. In addition, He is a technical advisor of J-Tech Corporation in Japan. His research interests include Network Security, Group Communication Security, Protocol Security, Information Leakage, Spam Mail Prevention, Web Spam Detection, group authentication, and Cryptography.